\documentclass[12pt]{amsart} 

\newcommand{\kahler}{K\"ahler\ }
\newcommand{\sqrtn}{\sqrt{N}}
\newcommand{\wt}{\widetilde}
\newcommand{\wh}{\widehat}
\newcommand{\PP}{{\mathbb P}}

\newcommand{\C}{{\mathbb C}}

\newcommand{\CP}{\C\PP}
\renewcommand{\d}{\partial}
\newcommand{\dbar}{\bar\partial}
\newcommand{\ddbar}{\partial\dbar}

\newcommand{\E}{{\mathbf E}\,}

\newcommand{\half}{{\frac{1}{2}}}
\newcommand{\vol}{{\operatorname{Vol}}}

\newcommand{\SU}{{\operatorname{SU}}}
\newcommand{\FS}{{{\operatorname{FS}}}}

\renewcommand{\phi}{\varphi}

\newcommand{\ccal}{\mathcal{C}}
\newcommand{\dcal}{\mathcal{D}}

\newcommand{\gcal}{\mathcal{G}}
\newcommand{\lcal}{\mathcal{L}}

\newcommand{\ocal}{\mathcal{O}}

\newcommand{\al}{\alpha}

\newcommand{\la}{\lambda}

\newcommand{\de}{\delta}

\newcommand{\om}{\omega}

\newtheorem{theo}{{\sc Theorem}}[section]

\newtheorem{lem}[theo]{{\sc Lemma}}

\newenvironment{rem}{\medskip\noindent{\it Remark:\/} }{\medskip}

\title[Poincar\'e-Lelong approach to
correlations between zeros]
{Poincar\'e-Lelong approach to universality and scaling of
correlations between zeros}

\author{Pavel Bleher}
\address{Department of Mathematical Sciences, IUPUI, Indianapolis, IN
46202,
USA}
\email{bleher@math.iupui.edu}
\author{Bernard Shiffman}
\address{Department of Mathematics, Johns Hopkins University, Baltimore,
MD
21218, USA} 
\email{shiffman@math.jhu.edu}
\author{Steve Zelditch}
\address{Department of Mathematics, Johns Hopkins University, Baltimore,
MD
21218, USA} 
\email{zel@math.jhu.edu}

\thanks{Research partially supported by NSF grants \#DMS-9623214 
(first author), \#DMS-9800479  (second author),
\#DMS-9703775 (third author).} 

\date{March 4, 1999}

\begin{document}

\begin{abstract}  This note is concerned with the scaling limit as $N
\to \infty$ of  $n$-point correlations between zeros of random
holomorphic polynomials of degree $N$ in
$m$ variables. More generally we study correlations between zeros of
holomorphic sections of powers
$L^N$ of any positive holomorphic line bundle $L$ over a compact \kahler
manifold. Distances are rescaled so that the average density of zeros is
independent of $N$. Our main result is that the scaling limits of the 
correlation functions and, more generally, of the ``correlation forms"
are universal, i.e. independent of the bundle $L$, manifold $M$ or point on
$M$.   
\end{abstract}

\maketitle

\section*{Introduction} 
This note is a companion to our article \cite{BSZ}, in which we study
the correlations between the zeros of a random holomorphic section $s \in
H^0(M, L^N)$ of a power $L^N$ of a positive line bundle $L \to M$ over a
compact
$m$-dimensional complex manifold $M$.  Since the hypersurface volume of the
zeros of a section of $L^N$ in a ball $U$ around a given point $z_0$ is
$\sim N {\rm Vol\,}(U)$, we rescale $U \to \sqrt{N} U$ to get a density of
zeros independent of $N$. After expanding $U$ this way, all manifolds and
line bundles appear asymptotically alike, and it is natural to ask if the
{\it local statistics} of zeros are universal, i.e.  independent of $L, M,
\omega$ and $z_0$. To define our statistics, we first provide $H^0(M,L^N)$
with a natural Gaussian measure (see \S\S\ref{cxgeom}--\ref{random}).  The
local statistics are measured by the scaled {\it $n$-point zero correlation
forms}
$\vec K_{n}^N(\frac{z^1}{\sqrtn},\dots,\frac{z^n}{\sqrtn})$, $z^j\in
\sqrtn\,U$ (see \S\ref{CC}). They are smooth forms on the ``off-diagonal"
domain 
$\gcal^m_n\subset C^{mn}$ consisting of $n$-tuples of distinct points $z^j\in
\C^m$, and their norms define scaled zero correlation measures
$\wt K_{n}^N(\frac{z^1}{\sqrtn},\dots,\frac{z^n}{\sqrtn})$. (The correlation
forms extend to all of $\C^{mn}$ as currents of order
0, and hence the same holds for the correlation measures.) In
\cite{BSZ}, we used geometric probability methods and a (universal) scaled
Szeg\"o kernel to prove that there exist universal limits as $N \to \infty$
of these  correlation measures and more generally of the correlations
between simultaneous zeros of
$k\le m$ sections. Here we use a complex analytic approach based on the
Poincar\'e-Lelong formula for the currents of integration over the zero set
of a section, together with the scaled Szeg\"o kernel from
\cite{BSZ}, to give a proof of universality for the correlation forms. 
This  approach, although limited to the hypersurface case, allows for a
result on the level of forms and a somewhat simpler proof.  

Our universality theorem is as follows:

\bigskip

\pagebreak
\noindent{\sc Main Theorem.} {\it There is a universal current $\vec
K_{n}^{\infty}\in\dcal'{}^{(m-1)n,(m-1)n}(\C^{mn})$ such that the following
holds: suppose that $(L,h)$ is a positive Hermitian line bundle on an
$m$-dimensional compact complex manifold $M$, and let $\vec K_{n}^N$ be the
$n$-point zero correlation current on $M^n$. Suppose $z^0\in M$ and choose
local holomorphic coordinates in $M$ about $z^0$ such that $\Theta_h|_{z^0}
=\ddbar|z|^2$. Then
$$\vec K_{n}^N\left(\frac{z^1}{\sqrt{N}}, \dots,\frac{z^n}{\sqrt{N}}\right)
=\vec K_{n}^{\infty}(z^1,\dots,z^n)+ O\left(\frac{1}{\sqrtn}\right)\,.$$
Furthermore, $\vec
K_{n}^{\infty}$ is a smooth form on the off-diagonal
domain 
$\gcal^m_n$, and the error term has $k^{th}$ order derivatives $\le
\frac{C_{A,k}}{\sqrtn}$ on each compact subset $A\subset \gcal^m_n$,
$\forall k\ge 0$. 
}

\bigskip Our method leads to integral formulae for these universal limit
forms, although the details rapidly become complicated as the number $n$ of
points increase.  For the case $m=2$, we carry out the calculation in
complete detail in dimension one and also use the method to obtain an
explicit formula for the scaling limit pair correlation measures in all
dimensions (Theorems \ref{Hformula} and \ref{mformula}). In particular, our
formula  gives the scaling limit pair correlations for
$\SU(m+1)$-polynomials (which are sections of powers of the hyperplane bundle
over complex projective space $\CP^m$).  The universal formula in
dimension one agrees, as it must, with that of Bogomolny-Bohigas-Leboeuf
\cite{BBL} and Hannay \cite{H} in the case of random $\SU(2)$-polynomials.
Similar formulas for correlations of zeros of real polynomials
were given in \cite{BD}. 

Before we get started on the proof,   a few heuristic remarks on
correlation  measures and forms may be helpful. Roughly speaking, 
$\wt K_{n}^N(z^1, \dots,z^n)$ gives the conditional probability density of
the zero divisor of a random section $s$ (simultaneously) intersecting small
balls around $z^{k+1}, \dots, z^n$, given  that the zero divisor
(simultaneously) intersects small balls around $z^1,
\dots, z^k$. The correlation form $\vec K_{n}^N$ gives a more refined
conditional probability:  Let $Y$ denote the  set of holomorphic tangent
hyperplanes in $M$.  (We can identify $Y$ with the projectivized holomorphic
cotangent bundle of
$M$.) Then $\vec K_{n}^N$ gives the conditional probability  that the zero
divisor has tangent hyperplanes in small balls in $Y$ above
$z^{k+1}, \dots, z^n$, given that it has  tangents in small balls above $z^1,
\dots, z^k$.

\medskip

\noindent{\bf Acknowledgment.} The first draft of this paper was
completed while the
third author was visiting the Erwin Schrodinger Institute in July 1998.
He wishes to thank that institution
for its hospitality and financial support.

\section{Notation}\label{notation}

We summarize here the notation from complex analysis that we will need in the
proof.  This notation is the same as in  \cite{SZ} and \cite{BSZ}, except
that different normalizations for the metric and volume form are used in
\cite{SZ}.

\subsection{Complex geometry}\label{cxgeom}

We denote by $(L, h) \to M$ a holomorphic line bundle with  smooth
Hermitian metric $h$ whose
curvature form
\begin{equation}\label{curvature}\Theta_h=-\ddbar
\log\|e_L\|_h^2\;,\end{equation} 
is a positive (1,1)-form.  Here,  $e_L$ is a local non-vanishing
holomorphic section of $L$ over an open set $U\subset M$, and
$\|e_L\|_h=h(e_L,e_L)^{1/2}$ is the $h$-norm of $e_L$.  As in
\cite{BSZ},
we give $M$ the
Hermitian metric corresponding to the \kahler form
$\omega=\frac{\sqrt{-1}}{2}\Theta_h$ and the induced Riemannian volume
form 
\begin{equation}\label{dV} dV_M= \frac{1}{m!}
\omega^m\;.\end{equation}

We denote by $H^0(M, L^{N})$ the space of  holomorphic
sections of
$L^N=L\otimes\cdots\otimes L$.   The metric $h$ induces Hermitian
metrics
$h^N$ on $L^N$ given by $\|s^{\otimes N}\|_{h^N}=\|s\|_h^N$.  We give
$H^0(M,L^N)$ the Hermitian inner product
\begin{equation}\label{inner}\langle s_1, s_2 \rangle = \int_M h^N(s_1,
s_2)dV_M \quad\quad (s_1, s_2 \in H^0(M,L^N)\,)\;,\end{equation} and we
write
$|s|=\langle s,s \rangle^{1/2}$.  

For a holomorphic section $s\in H^0(M, L^N)$, we let $Z_s$ denote the
current
of integration over the zero divisor of $s$:
$$(Z_s,\phi)=\int_{Z_s}\phi\,,\quad \phi\in\dcal^{m-1,m-1}(M)\,.$$ The
Poincar\'e-Lelong formula (see e.g., \cite{GH}) expresses the
integration
current of a holomorphic section $s=ge_L^{\otimes N}$ in the form:
\begin{equation}
\label{PL} Z_s  = \frac{i}{\pi}
\ddbar \log |g|
= \frac{i}{\pi}
\ddbar \log \|s\|_{h^N} + N\omega\;.\end{equation} We also denote
by $|Z_s|$
the Riemannian $(2m-2)$-volume along the regular points of $Z_s$,
regarded as
a measure on $M$:
\begin{equation}\label{volmeasure} (|Z_s|,\phi)=\int_{Z_s^{\rm reg}}\phi
d\vol_{2m-2}=\frac{1}{(m-1)!}\int_{Z_s^{\rm reg}}\phi
\omega^{m-1}\,;\end{equation}  i.e., 
$|Z_s|$ is the total variation measure of the current of integration
over
$Z_s$: 
\begin{equation}\label{volmeasure2}|Z_s|=
Z_s\wedge{\textstyle \frac{1}{(m-1)!}}\omega^{m-1}\,.\end{equation}

\subsection{Random sections and Gaussian measures}\label{random}

We now  give $H^0(M,L^N)$ the complex
Gaussian probability measure
\begin{equation}\label{gaussian}d\mu(s)=\frac{1}{\pi^{d_N}}e^{-
|c|^2}dc\,,\qquad s=\sum_{j=1}^{d_N}c_jS_j^N\,,\end{equation} where
$\{S_j^N\}$ is an orthonormal basis for
$H^0(M,L^N)$ and $dc$ is $2d_N$-dimensional Lebesgue measure.  This
Gaussian
is characterized by the property that the $2d_N$ real variables $\Re
c_j,
\Im c_j$ ($j=1,\dots,d_N$) are independent random variables with mean 0
and
variance $\half$; i.e.,
$$\E c_j = 0,\quad \E c_j c_k = 0,\quad  \E c_j \bar c_k = \de_{jk}\,.$$
Here and throughout this paper, $\E$ denotes expectation: $\E\phi = \int
\phi d\mu$.

We then  regard the currents $Z_s$ (resp.\  measures $|Z_s|$), as
current-valued (resp.\  measure-valued) random variables on the
probability space $(H^0(M,L^N), d\mu)$; i.e., for each test form
(resp.\ function) $\phi$,
$(Z_s,\phi)$ (resp.\  $(|Z_s|,\phi)$) is a complex-valued random
variable.

Since the zero current $Z_s$ is unchanged when $s$ is multiplied by an
element of $\C^*$, our results are the same if we instead regard $Z_s$ as a
random variable on the unit sphere $SH^0(M,L^N)$ with Haar probability
measure.  We prefer to use Gaussian measures in order to facilitate
our computations.

\subsection{Correlation currents}\label{CC}

The $n$-point correlation current of the zeros is  the current on $M^n =
M
\times M \times \cdots
\times M$ ($n$ times) given by 
\begin{equation} \vec K_n^N(z^1, \dots, z^n):= \E(Z_s(z^1) \otimes
Z_s(z^2)
\otimes \dots \otimes
Z_s(z^n)) \end{equation}
in the sense that for any test form $\phi_1(z^1) \otimes \dots \otimes
\phi_n(z^n) \in \dcal^{m-1,m-1}(M) \otimes \dots
\otimes \dcal^{m-1,m-1}(M)$,
\begin{equation} \big(\vec K_n^N(z^1, \dots, z^n),
\phi_1(z^1) \otimes \dots \otimes \phi_n(z^n)\big)
= \E \left[\big( Z_s,\phi_1\big)\big( Z_s,\phi_2\big)\cdots 
\big(Z_s, \phi_n\big)\right].
\end{equation}
In a similar way we define the $n$-point correlation measures $\wt K_n^N$ as
the ``total variation measures" of
the $n$-point correlation currents:
\begin{equation}\wt K_n^N(z^1, \dots, z^n)=\vec K_n^N(z^1, \dots, z^n)\wedge
\frac{1}{(m-1)!}\om_{z^1}^{m-1}\wedge\cdots\wedge 
\frac{1}{(m-1)!}\om_{z^n}^{m-1}
\,,\end{equation} i.e.,
\begin{equation}\big(\wt K_n^N(z^1, \dots, z^n),
\phi_1(z^1) \dots  \phi_n(z^n)\big)
= \E \big[(|Z_s|, \phi_1)(|Z_s|, \phi_2)\cdots
(|Z_s|, \phi_n)\big] \end{equation}
where $\phi_j \in \ccal^0(M).$

\begin{rem} In the case of pair correlation on a Riemann surface ($n=2, \dim
M=1$), the correlation measures take the form
$$ \vec K^N_2(z,w) = [\Delta]\wedge (\vec K^N_1(z)
\otimes 1)  + \kappa^N(z,w)\omega_z\otimes
\omega_w \quad (N\gg 0)$$ where $[\Delta]$ denotes the current of
integration along the diagonal $\Delta=\{(z,z)\}\subset M\times M$, and
$\kappa^N\in\ccal^\infty(M\times M)$. \end{rem}

\subsection{Szego kernels}

As in \cite{Z,SZ,BSZ} and elsewhere, we analyze the
$N \to \infty$ limit by lifting it to a principal $S^1$ bundle $\pi: X
\to M$. Let us recall how this goes.

We denote by $L^*$ the dual line bundle to $L$, and define $X$ as the circle
bundle $X=\{\la \in L^* : \|\la\|_{h^*}= 1\}$, where $h^*$ is the norm on
$L^*$ dual to $h$. We can view  $X$ as the boundary of the disc bundle $D =
\{\la
\in L^* : \rho(\la)>0\}$, where $\rho(\la)=1-\|\la\|^2_{h^*}$. The disc
bundle
$D$ is strictly pseudoconvex in $L^*$, since $\Theta_h$ is positive, and
hence
$X$ inherits the structure of a strictly pseudoconvex CR manifold.
Associated to
$X$ is the contact form $\al= -i\partial\rho|_X=i\dbar\rho|_X$.  We also give
$X$ the volume form
\begin{equation}\label{dvx}dV_X=\frac{1}{m!}\al\wedge 
(d\al)^m=\al\wedge\pi^*dV_M\,.\end{equation}

The setting for our analysis of the Szeg\"o kernel is the Hardy space
$H^2(X)
\subset \lcal^2(X)$ of square integrable CR functions on $X$, where we use
the inner product
\begin{equation}\label{unitary} \langle  F_1, F_2\rangle
=\frac{1}{2\pi}\int_X F_1\overline{F_2}dV_X\,,\quad
F_1,F_2\in\lcal^2(X)\,.\end{equation} We let $r_{\theta}x =e^{i\theta} x$
($x\in X$) denote the
$S^1$ action on $X$. The action $r_\theta$ commutes with the Cauchy-Riemann
operator
$\bar{\partial}_b$; hence $H^2(X) = \bigoplus_{N = 0}^{\infty} H^2_N(X)$,
where
$$H^2_N(X) = \{ F \in H^2(X): F(r_{\theta}x) = e^{i N \theta} F(x)
\}\,.$$ A section $s_N$ of $L^{N}$ determines an equivariant function
$\hat{s}_N$ on $X$:
\begin{equation}\label{snhat}\hat{s}_N(z, \lambda) = \left( \lambda^{\otimes
N}, s_N(z)
\right)\,,\quad (z,\la)\in X\,;\end{equation} then $\hat s_N(r_\theta x) =
e^{iN\theta} s_N(x)$.  The map $s\mapsto
\hat{s}$ is a unitary equivalence between $H^0(M, L^{\otimes N})$ and
$H^2_N(X)$. 

We let $\Pi_N : \lcal^2(X) \rightarrow H^2_N(X)$ denote the orthogonal
projection.  The Szeg\"o kernel $\Pi_N(x,y)$ is defined by
\begin{equation} \Pi_N F(x) = \int_X \Pi_N(x,y) F(y) dV_X (y)\,,
\quad F\in\lcal^2(X)\,.
\end{equation} It can be given as
\begin{equation}\label{Szego}\Pi_N(x,y)=\sum_{j=1}^{d_N}\wh 
S_j^N(x)\overline{\wh S_j^N(y)}\,,\end{equation} where
$S_1^N,\dots,S_{d_N}^N$ form an orthonormal basis of $H^0(M,L^N)$.  

\section{Scaling}

In order that we may study the local nature of the random variable
$Z_s$, we
fix a point $z^0\in M$ and choose a holomorphic coordinate chart
$\Psi:\Omega,0\to U,z_0$ ($\Omega\subset \C^m,\ U\subset M$) such that
\begin{equation}\label{coord} \Psi^*\omega_{z^0} 
=\left.\frac{i}{ 2}\sum_{j=1}^m dz_j\wedge
d\bar z_j\right|_0\;.\end{equation}
For example, if $L$ is the hyperplane section bundle $\ocal(1)$ over
$\C\PP^m$
with the Fubini-Study metric $h_\FS$, and
$z_0=(1:0:\dots:0)$, then the coordinate chart $$\Psi:\C^m\to
U=\{w\in\CP^m:w_0\neq 0\}\,,\quad \Psi(z)=(1:z_1:\dots :z_m)$$ (i.e.,
$z_j=w_j/w_0$) satisfies (\ref{coord}).

To simplify notation, we identify $U$ with $\Omega$.  For a current
$T\in\dcal'{}^{p,q}(\Omega)$, we write
$$T\left(\frac{z}{\sqrtn}\right)=\left(\tau_{\sqrtn}\right)_* T\in
\dcal'{}^{p,q}(\sqrtn\Omega)\qquad (\tau_\la(z)=\la z)\,.$$ (In particular,
if
$T=\sum T_{jk}(z)dz_j\wedge d\bar z_k$, then $T(\frac{z}{\sqrtn})=
\frac{1}{N}\sum T_{jk}(\frac{z}{\sqrtn})dz_j\wedge d\bar z_k$.)

We define the {\it rescaled zero current} of
$s \in
H^0(M,L^N)$  by
\begin{equation} \label{zhat} \wh
Z^N_s(z):=Z_{s}\left(\frac{z}{\sqrtn}\right)\,.  \end{equation} 
The scaled $n$-point correlation currents
are then  defined by:
\begin{equation}\label{def-var} \E\left( \wh{Z}^N_s(z^1) \otimes 
\wh{Z}^N_s(z^2) \otimes \cdots \otimes \wh{Z}^N_s(z^n)\right)
=\vec
K_n^N\left(\frac{z^1}{\sqrtn},\dots,\frac{z^n}{\sqrtn}\right)\in
\dcal'{}^{n,n}(M^n).
\end{equation}

Following the approach of \cite{SZ}, we fix an orthonormal basis
$\{S^N_j\}$
of $H^0(M, L^{N})$ and write $S^N_j = f^N_j e_L^{\otimes N}$ over $U$.
Any
section in $H^0(M, L^{N})$ may then be written as $s = \sum_{j=1}^{d_N}
c_j
f^N_j e_L^N$. To simplify the notation we
let $f^N=(f^N_1,\ldots,f^N_{d_N}):U\to \C^{d_N}$  and we put $$
\sum_{j=1}^{d_N} c_j f_j = c\cdot f^N\;.$$ Hence
\begin{equation}\label{PL2} Z_s = \frac{\sqrt{-1}}{\pi} \ddbar
\log| c\cdot f^N|\;,\quad \wh Z^N_s
=\frac{\sqrt{-1}}{
\pi}\d_z\dbar_z\log\big|c\cdot f^N\big(\frac{z}{\sqrtn}\big)
\big|
\end{equation}  and therefore
\begin{equation} \wh
Z_s(z^1) \otimes\cdots\otimes \wh Z_s(z^n)= \left(\frac{i }{
\pi}\right)^n\d_{z^1}\dbar_{z^1}\cdots\d_{z^n}\dbar_{z^n}
\left[\log|c\cdot
f^N(\frac{z^1}{\sqrtn})|\cdots\log| c\cdot
f^N(\frac{z^n}{\sqrtn})|\right]\;. \end{equation}
We then can write the rescaled correlation currents in the form
\begin{equation}\label{PLcorscaled}\begin{array}{l}\displaystyle  
\vec K^N_n\left(\frac{z^1}{\sqrtn},\dots,\frac{z^n}{\sqrtn}\right)=
\E\big(\wh Z_s(z^1) \otimes\cdots\otimes \wh  
Z_s(z^n)\big)\\[12pt] \displaystyle\quad = \left(\frac{i }{
\pi}\right)^n  \d_{z^1}\dbar_{z^1}\cdots\d_{z^n}\dbar_{z^n}
\int_{\C^{d_N}}
\log\left|c\cdot
f^N\left(\frac{z^1}{\sqrtn}\right)\right| \cdots\log\left|c\cdot
f^N\left(\frac{z^n}{\sqrtn}\right) 
\right|\frac{e^{-|c|^2}}{\pi^{d_N}}
dc\;.\end{array}
\end{equation}

\subsection{Scaling limit of the Szego kernel}\label{scalingszego}

The asymptotics of the Szeg\"o kernel along the diagonal were given by
\cite{Ti} and \cite{Z}:
\begin{equation}\label{diag}\frac{\pi^m}{N^m}\Pi_N(x,x)= 1
+O(N^{-1})\,.\end{equation}  For our proof of the Main Theorem,
we need the following lemma from \cite{BSZ}, which gives the
`near-diagonal' asymptotics of the Szego kernel.
 
\begin{lem} \label{neardiag}  Let $z^0\in M$ and choose local coordinates
$\{z^j\}$ in a neighborhood of
$z_0$ so that $z^0=0$ and $\Theta_h(z_0)=\sum dz^j\wedge d\bar z^j$. Then
\begin{eqnarray*}\frac{\pi^m}{N^m}\Pi_N(\frac{z}{\sqrtn},\frac{\theta}{N };
\frac{w}{\sqrtn},\frac{\phi}{N})&=&e^{i2\pi (\theta-\phi)+i\Im (z\cdot \bar
w)-\half |z-w|^2} + O(N^{-1/2})\;.\end{eqnarray*}
\end{lem} Here, $(z,\theta)$ denotes the point
$e^{i\theta}\|e_L(z)\|_he^*_L(z)\in X$, and similarly for $(w,\phi)$.  In
(\ref{diag}) and Lemma \ref{neardiag}, the expression 
$O(N^\al)$ means a term with $k^{\rm th}$ order derivatives $\le C_k 
N^\al$, for all $k\ge 0$. Lemma \ref{neardiag} says that the Szeg\"o kernel
has a universal scaling limit.  In fact, its scaling limit is the first
Szeg\"o kernel of the reduced Heisenberg group; see \cite{BSZ}.

\section{Universality}\label{universality}

All the ideas of the proof of the Main Theorem occur in the
simplest case  $n = 2$.
So first we prove universality in that case and then extend the proof to
general $n$. 

Thus, our first object is to prove that the  large $N$ limit of the
rescaled pair
correlation current (from (\ref{PLcorscaled}) with $n=2$)
\begin{equation}\label{PLcor2}\begin{array}{l}\displaystyle \vec
K^N_2\left(\frac{z}{\sqrtn},\frac{w}{\sqrtn}\right)=
\E\left(\wh Z^N_s(z) \otimes \wh Z^N_s(w)\right)\\[10pt]
\displaystyle\quad\quad\quad=\frac{-1 }{ \pi^2}
\d_{z}\dbar_{z}\d_w\dbar_w
\int_{\C^{d_N}}\log\left|c\cdot
f^N(\frac{z}{\sqrtn})\right|\,\log\left| c\cdot
f^N(\frac{w}{\sqrtn})\right| \frac{e^{-|c|^2}}{\pi^{d_N}} dc \end{array}
\end{equation}
is universal.  

As in \cite{SZ}, we write $f^N=|f^N|u^N$ and expand the integrand in
(\ref{PLcor2}): \begin{eqnarray}\log|c\cdot
f^N(\frac{z}{\sqrtn})|\log|c\cdot
f^N(\frac{w}{\sqrtn})|&=&
\log |f^N(  \frac{z}{\sqrt{N}})| \log |f^N(
\frac{w}{\sqrt{N}})|\nonumber \\
&&+\log|f^N(\frac{z}{\sqrt{N}})| \log |c\cdot u^N(\frac{w}{\sqrt{N}})
|\nonumber \\&&+ \log |f^N(\frac{w}{\sqrt{N}})| \log |c\cdot
u^N(
\frac{z}{\sqrt{N}}) |\nonumber \\ && +\log |c\cdot
u^N(\frac{z}{\sqrt{N}})|  \log |c\cdot
u^N(\frac{w}{\sqrt{N}})
|\;.\label{expand}\end{eqnarray} 
Let us denote the terms resulting from this expansion by $E_1,\ E_2,\
E_3,\
E_4$, respectively. In particular, 
\begin{equation}\label{E1}E_1=\frac{-1 }{
\pi^2}\d_z\dbar_z\d_w\dbar_w \left[\log \big|f^N( \frac{z}{\sqrt{N}},
\frac{z}{\sqrt{N}})\big| \log \big|f^N( \frac{w}{\sqrt{N}},
\frac{w}{\sqrt{N}})\big|
\right]\;.\end{equation} 

By (\ref{snhat}), $\wh S^N_j(z,\theta)=e^{iN\theta} \|e_L(z)\|_h^{N} f^N_j
(z)$, where $(z,\theta)$ are the
coordinates in $X$ given in \S \ref{scalingszego}. By (\ref{Szego}),
\begin{equation}\label{szego1}
\Pi_N(z,w)=\|e_L(z)\|_h^{N} \|e_L(w)\|_h^{N}\langle f^N(z), f^N(w)\rangle\,,
\end{equation} where we write $\Pi_N(z,w)=\Pi_N(z,0;w,0)$.
Since $\Pi_N(z,z)^{1/2} = \|e_L(z)\|^N_h |f^N(z)|$, each
factor in (\ref{E1})
has the form $\half\log \Pi_N(\frac{z}{\sqrt{N}}, \frac{z}{\sqrt{N}}) - N
\log
\|e_L(\frac{z}{\sqrt{N}})\|_h$. By
(\ref{diag}), $\log
\Pi_N(\frac{z}{\sqrt{N}}, \frac{z}{\sqrt{N}})\to 0$ as $N\to\infty$. On the
other hand
$$-iN\d_z\dbar_z \log
\|e_L(\frac{z}{\sqrt{N}})\|_h = \omega (\frac{z}{\sqrt{N}})\;.$$ Hence the
first term converges to the normalized Euclidean (double) \kahler form:
\begin{equation}\label{E1limit} E_1=\frac{i}{ 2\pi}\ddbar|z|^2 \wedge
\frac{i}{
2\pi}\ddbar|w|^2 +O(\frac{1}{N})\;.\end{equation}

The middle two terms vanish since
the integrals in $E_2$ and $E_3$ are independent of $w$ and $z$
respectively (see \cite[\S 3.2]{SZ}). The ``interesting term'' is
therefore
\begin{equation}\label{interesting} E_4= \frac{-1
}{\pi^2}\d_z\dbar_z\d_w\dbar_w  \int_{\C^{d_N}}\log|
c\cdot u^N(\frac{z}{\sqrtn})|\log|c\cdot
u^N(\frac{w}{\sqrtn})| \frac{e^{-|c|^2}}{\pi^{d_N}} dc
\; .\end{equation} To evaluate $E_4$, we consider
the integral \begin{equation}\label{GN} G_2^N(x^1,x^2) := \int_{\C^{d_N}}
\log
|c\cdot x^1| \log | c\cdot x^2| \frac{e^{-|c|^2}}{\pi^{d_N}} dc\quad
(x^1, x^2\in
\C^{d_N}) \end{equation}
with $x^1 =  u^N(\frac{z}{\sqrtn}),\ x^2 =  u^N(\frac{w}{\sqrtn})$. 
To simplify it, we construct a Hermitian orthonormal basis $\{e_1,
\dots, e_{d_N}\}$ for $\C^{d_N}$ such that $x^1=e_1$ and
\begin{equation}\label{2D}
x^2 = \xi_{1} e_1 + \xi_{2} e_2,\quad\xi_{1} = \langle x^2,
x^1\rangle,\;\; \xi_{2} = \sqrt{1 - |\xi_{1}|^2}.
\end{equation}
This is possible because we can always multiply $e_2$ by a phase $e^{i
\theta}$ so that $\xi_{2}$ is positive real. We then make a unitary
change of variables to express
the integral in the  $\{e_j\}$ coordinates. Since the Gaussian is
$U(d_N)$-invariant, (\ref{GN}) simplifies to
\begin{equation} \label{2dim} G_2^N(x^1,x^2) = G_2(\xi_1,\xi_2)=
\frac{1}{\pi^2} \int_{\C^2 }e^{-(|c_1|^2 + |c_2|^2)}  \log |\xi_{1}|\log
|c_1\xi_{1} + c_2 \xi_{2} | dc_1 dc_2\end{equation}
(where we used the fact that  the  Gaussian integral in
each
$c_j, j \geq 3$ equals one by construction).
By performing a rotation of the $c_1$ variable, we may replace
$\xi_1$ with $|\xi_1|$ and replace $G_2(\xi_1,\xi_2)$ with
\begin{equation}\label{Gcos}G(\cos\theta):=G_2(\cos \theta, \sin\theta)
\,,\end{equation}
where $\cos\theta = |\xi_1|=|\langle x^1,x^2\rangle|$,
$0\leq\theta\leq\pi/2$. Hence (\ref{interesting}) becomes
\begin{equation}\label{interesting-simp} E_4=
\frac{-1 }{\pi^2}\d_z\dbar_z\d_w\dbar_w G(\cos\theta_N)\;,\quad
\cos\theta_N= 
\big|\big\langle u^N(\frac{z}{\sqrt{N}}),
u^N(\frac{w}{\sqrt{N}})\big\rangle\big|\;.\end{equation}

By the universal scaling formula for the Szego kernel
(Lemma~\ref{neardiag}) and (\ref{szego1}),
we have
\begin{equation}\label{thetalimit}\cos\theta_N 
=\frac{|\Pi_N(z,w)|}{\Pi_N(z,z)^{1/2}\Pi_N(w,w)^{1/2}}
=  e^{-\half
|z-w|^2} + O(N^{-\half})\;.\end{equation}
Thus we get the universal formula:
\begin{equation}\label{UPCC} \vec K_2^{\infty}(z,w) = \frac{i}{
2\pi}\ddbar|z|^2 \wedge \frac{i}{
2\pi}\ddbar|w|^2 +   \frac{-1 }{\pi^2}\d_z\dbar_z\d_w\dbar_w G(e^{-\half
|z - w|^2}). \end{equation}
This completes the proof for the pair correlation case $n=2$. (Notice that 
the formula has the same form in all dimensions.)   

The proof for general $n$ is similar. We again write $f^N=|f^N|u^N$ and
expand the integrand in (\ref{PLcorscaled}):
$$ \begin{array}{l} \log|c\cdot
f^N(\frac{z^1}{\sqrtn})|\log|c\cdot
f^N(\frac{z^2}{\sqrtn})|\cdots \log|c\cdot
f^N(\frac{z^n}{\sqrtn})| \\ \\ 
= \log |f^N(  \frac{z^1}{\sqrt{N}})| \log |f^N( \frac{z^2}{\sqrt{N}})|
\cdots \log|f^N(\frac{z^n}{\sqrtn})|
\nonumber \\ \\ 
+\log|f^N(\frac{z^1}{\sqrt{N}})|\log |f^N( \frac{z^2}{\sqrt{N}})| \cdots
\log|f^N(\frac{z^{n-1}}{\sqrtn})| \log |c\cdot
u^N(\frac{z^n}{\sqrt{N}})
|\nonumber \\ \\  + \cdots \\ \\+\log |c\cdot
u^N(\frac{z^1}{\sqrt{N}})|  \log |c\cdot
u^N(\frac{z^2}{\sqrt{N}})| \cdots \log |c\cdot
u^N(\frac{z^n}{\sqrt{N}}) |\;.\end{array}$$
We denote the terms resulting from this expansion by $E_1,\dots, E_{2^n}$,
respectively. As before, the first term converges to the normalized Euclidean
``$n$-fold" \kahler form:
$$E_1=\frac{i}{2\pi}\ddbar|z^1|^2\wedge \cdots
\wedge \frac{i}{2\pi}\ddbar|z^n|^2 +O(\frac{1}{N})\,.$$  The $E_{2^n}$
term is obtained from the function
\begin{equation}\label{GNn} G_n^N(x^1, x^2, \dots, x^n) :=
\int_{\C^{d_N}} \log
|c\cdot x^1| \log |c\cdot x^2|\cdots \log
|c\cdot x^n|  \frac{e^{-|c|^2}}{\pi^{d_N}} dc\,, \end{equation} 
$x^1,x^2, \dots, x^n\in
\C^{d_N}$.  Precisely, we substitute
\begin{equation}\label{xj}  x^j = u^N(\frac{z^j}{\sqrt{N}})\end{equation}
in (\ref{GNn}) and apply the operator $\left(\frac{i }{
\pi}\right)^n  \d_{z^1}\dbar_{z^1}\cdots\d_{z^n}\dbar_{z^n}$.
 As above,  we  define  a special
Hermitian orthonormal basis
$\{e_1,
\dots,
e_n\}$ for the n-dimensional complex subspace spanned by $\{x_1, \dots,
x_n\}.$  We put:
$$\begin{array}{ll} x^1 =  e_1 \\
x^2 = \xi_{21} e_1 + \xi_{22}e_2 &  \xi_{22} =
\sqrt{1 - | \xi_{21}|^2} \\ \vdots \\
x^n = \xi_{n1} e_1 + \dots + \xi_{nn} e_n\qquad & \xi_{nn} = \sqrt{1 -
\sum_{j \leq n-1}
|\xi_{nj}|^2 }. \end{array}$$
Such a basis exists because we can always multiply $e_j$ by a 
phase $e^{i \theta}$ so that the last component $\xi_{jj}$ is positive
real. 
We complete $\{e_j\}$ to a basis
of $\C^{d_N}$, and we now let $c_j$ denote coordinates relative to this
basis. As above, we rewrite the Gaussian integral in these coordinates.
After integrating out the variables $\{c_{n+1}, \dots,
c_{d_N}\}$, (\ref{GNn}) simplifies to
the $n$-dimensional complex Gaussian integral
\begin{equation}\label{ndim}\begin{array}{lll} G^N_n(x^1,\dots,x^n)&=&
G_n(\xi_{21},
\xi_{22},
\dots,
\xi_{nn})\\[10pt] &=&
\frac{1}{\pi^n} \int_{\C^n} e^{-|c|^2} \log|c_1| \log
|c_1 \xi_{21} +
c_2 \xi_{22}|\cdots \log |c_1 \xi_{n1} + 
\dots c_n \xi_{nn}| dc\,.\end{array}\end{equation}
Note that the variables $\xi_{jk}$ depend on $N$; we write $\xi_{jk} =
\xi_{jk}^N$ when we need to indicate this dependence. 

To prove universality, we observe that the $\xi_{jk}$
are universal algebraic functions of the inner products $\langle x^a,
x^b \rangle$. Indeed,
\begin{equation}\label{lex}
\xi_{j1}\bar\xi_{k1}+\cdots+\xi_{jk}\bar\xi_{kk}=\langle
x^j,x^k\rangle\,,\quad 1\le k\le j\le n\,,\end{equation}
where we set $\xi_{11}=1$.  These algebraic functions are obtained by
induction (lexicographically) using (\ref{lex}). (The triangular matrix
$(\xi_{jk})$ is just the inverse of the matrix describing the Gram-Schmidt
process.)

By (\ref{xj}),  it follows that
the $\xi_{jk}^N$ 
are universal algebraic functions 
of the variables 
$$\left\langle
u^N({\textstyle \frac{z^j}{\sqrt{N}}}),
u^N({\textstyle\frac{z^k}{\sqrt{N}}})
\right\rangle=\frac{\Pi_N(\frac{z^j}{\sqrt{N}}, \frac{z^k}{\sqrt{N}})}{
\Pi_N(\frac{z^j}{\sqrt{N}},\frac{z^j}{\sqrt{N}})^{1/2}
\Pi_N(\frac{z^k}{\sqrt{N}},\frac{z^k}{\sqrt{N}})^{1/2}}
=e^{i\Im
(z^j\cdot \bar z^k)-\half |z^j-z^k|^2} + O({\textstyle\frac{1}{\sqrtn}})\,.$$
We note here that
\begin{equation}\label{det} |x^1\wedge\cdots\wedge x^n|^2=\det( \langle
x^j,x^k\rangle) \to \det \left(e^{i\Im (z^j\cdot \bar z^k)-\half
|z^j-z^k|^2}\right) =e^{-\sum|z^j|^2}\det\left(e^{z_j\cdot \bar
z_k}\right)\,.\end{equation} When the $z_j$ are distinct (i.e.,
$(z^1,\dots,z^n)\in \gcal^m_n$), the limit determinant in (\ref{det}) is
nonzero (see \cite{BSZ}) and thus $\xi_{jk}^N=
\xi_{jk}^\infty+O(\frac{1}{\sqrtn})$, where the $\xi_{jk}^\infty$ are
universal real-analytic functions of $z\in\gcal^m_n$.  We conclude that the 
$E_{2^n}$ term converges to a universal current:
$$E_{2^n}=\left(\frac{i}{\pi}\right)^n \d_{z^1}\dbar_{z^1}\cdots
\d_{z^n}\dbar_{z^n}G_n(\xi_{21}^\infty,\dots,\xi_{kk}^\infty) +
O({\textstyle\frac{1}{\sqrtn}})\,.$$

Consider now a general term $E_a$.  Suppose without loss of generality that
$E_a$ comes from 
$$ \log |c\cdot
u^N(\frac{z^1}{\sqrt{N}})|\cdots \log |c\cdot
u^N(\frac{z^k}{\sqrt{N}})|\log|f^N(\frac{z^{k+1}}{\sqrt{N}})|\cdots
\log|f^N(\frac{z^n}{\sqrtn})|\,.$$ As above we obtain
$$E_a=\left(\frac{i}{\pi}\right)^k \d_{z^1}\dbar_{z^1}\cdots
\d_{z^k}\dbar_{z^k}G_k(\xi_{21}^\infty,\dots,\xi_{kk}^\infty) 
\wedge \frac{i}{2\pi}
\ddbar |z^{k+1}|^2\wedge\cdots \wedge \frac{i}{2\pi}
\ddbar |z^{n}|^2 +O({\textstyle\frac{1}{\sqrtn}})\,.$$
Hence this
term also approaches a universal current. 
(As in the pair correlation case, terms with only one $u^N$ vanish.) \qed

\section{Explicit formulae}
 
We now calculate explicitly the limit pair correlation measures
$\wt K^{\infty}_2(z,w)$.

\subsection{Preliminaries}

The first step is to compute $\Delta
G(e^{-\half r^2})$, where $\Delta$ is the Euclidean Laplacian on
$\C^m$ and $r=|\zeta|$ ($\zeta\in\C^m$). To begin this computation, we
write
$a_j = r_j e^{i \phi_j}$ and then rewrite (\ref{2dim})--(\ref{Gcos}) as
\begin{equation} 
G(\cos\theta)=\frac{2}{\pi}\int_0^{\infty} \int_0^{\infty}\int_0^{2\pi}
r_1 r_2
e^{-(r_1^2 + r_2^2)}  \log r_1 \log |r_1 \cos\theta+ r_2 e^{i
\phi}\sin\theta| d\phi dr_1 dr_2\;.\end{equation}  We
now evaluate the inner integral by Jensen's formula, which gives
\begin{equation}  \int_0^{2\pi} \log|r_1 \cos \theta + r_2 \sin \theta e^{i
\phi}| d\phi = \left\{ \begin{array}{ll}2\pi \log (r_1 \cos \theta) &
\mbox{for}\;\;r_2 \sin \theta \leq r_1 \cos \theta \\ & \\ 2\pi\log (r_2
\sin \theta) & \mbox{for} \;\;r_2 \sin \theta\geq r_1 \cos \theta
\end{array}
\right. \end{equation} Hence \begin{equation} G(\cos\theta) = 4
\int_0^{\infty} \int_0^{\infty} r_1 r_2 e^{-(r_1^2 + r_2^2)}  \log r_1
\log
\max ( r_1 \cos \theta , r_2 \sin \theta)dr_1 dr_2. \end{equation}

Now change variables again with $r_1 = \rho \cos \phi, r_2 = \rho \sin
\phi$
to get \begin{equation} G(\cos\theta) = 4 \int_0^{\infty}
\int_0^{\pi/2} \rho^3 e^{-\rho^2} \log (\rho \cos \phi) \log\max (\rho
\cos \phi \cos \theta , \rho \sin \phi \sin \theta)   \cos\phi\sin \phi
d\phi d\rho\;. \end{equation} Since $$ \log \max( \rho \cos \phi \cos
\theta
, \rho \sin \phi \sin \theta) = \log (\rho \cos \phi \cos \theta) +
\log^+
(\tan \phi \tan \theta)\;,$$ we can write
$G=G_1+G_2$, where
\begin{eqnarray} \label{G1} G_1(\cos\theta)&=&4
\int_0^{\infty} \int_0^{\pi/2} \rho^3 e^{-\rho^2}   \log (\rho \cos
\phi)
\log ( \rho \cos \phi \cos \theta)  \cos\phi \sin \phi d\phi
d\rho\\ \label{G2} G_2(\cos\theta) &=&4\int_0^{\infty} \int_{\pi/2 -
\theta}^{\pi/2}\rho^3 e^{-\rho^2}   \log (\rho \cos \phi) \log ( \tan
\phi
\tan \theta) \cos\phi\sin \phi d\phi d\rho\;. \end{eqnarray}

{From} (\ref{G1}),
$G_1(\cos\theta)=C_1+C_2\log\cos\theta$ and thus $$G_1(e^{-\half r^2})=
C_1-\half C_2 r^2\;,$$
so that \begin{equation}\label{forgetG1}\Delta G_1(e^{-\half
r^2})=\left(\frac{d^2 }{ dr^2} +\frac{2m-1}{ r}\frac{d }{
dr}\right)(C_1-\half
C_2 r^2)=-2mC_2\;.\end{equation}

We now evaluate $\Delta G_2(e^{-\half r^2})$. Since the integrand in (\ref{G2})
vanishes when $\phi=\pi/2 -\theta$, we have
$$\frac{d }{ dr}G_2(\cos\theta)= 4 \left(\frac{d }{ dr}\log\tan\theta 
\right) \int_0^{\infty} \int_{\pi/2 -
\theta}^{\pi/2}\rho^3 e^{-\rho^2} \log (\rho \cos \phi) \cos\phi\sin
\phi
d\phi d\rho\;.$$  Substituting $\tan^2 \theta =
e^{r^2}-1$, we have  $$\frac{d }{ dr}\log\tan\theta =\frac{r}{
1-e^{-r^2}}\;.$$
Thus $$\frac{d }{ dr}G_2(e^{-\half r^2})=  \frac{4r}{
1-e^{-r^2}}(I_1+I_2)\;,$$ where
$$I_1=\int_0^{\infty} \int_{\pi/2 -
\theta}^{\pi/2}\rho^3 e^{-\rho^2} (\log \rho) \cos\phi\sin \phi
d\phi d\rho= C \sin^2 \theta = C(1-e^{-r^2})\;,$$
$$I_2=\int_0^{\infty} \int_{\pi/2 -
\theta}^{\pi/2}\rho^3 e^{-\rho^2} (\log \cos\phi) \cos\phi\sin \phi
d\phi d\rho\;.$$

We compute
\begin{eqnarray*}I_2 &=& \half\int_{\pi/2 -
\theta}^{\pi/2} (\log \cos\phi) \cos\phi\sin \phi
d\phi \ =\half\int_0^{\sin\theta}t\log tdt\\ 
&=&\frac{1}{ 8}(\sin^2\theta\log\sin^2\theta-\sin^2\theta)=\frac{1}{
8}(1-e^{-r^2})\left[\log (1-e^{-r^2})-1\right] \end {eqnarray*}
Thus \begin{equation} \label{d1} \frac{d }{ dr}G_2(e^{-\half r^2})=
\frac{r}{
2}\log (1-e^{-r^2}) +C'r\;.\end{equation}
Hence by (\ref{forgetG1}) and (\ref{d1}), \begin{eqnarray}\Delta
G(e^{-\half
r^2})&=& -2mC_2+ \left(\frac{d }{ dr}+\frac{2m-1}{
r}\right)\left(\frac{r}{
2}\log(1-e^{-r^2})+C'r\right)\nonumber \\&=&
m\log(1-e^{-r^2})+\frac{r^2}{
e^{r^2}-1} +C''\;.\label{d2}\end{eqnarray}

\subsection{Pair correlation in dimension 1} 

In dimension one, the pair correlation form is the same as the pair
correlation measure. We first give our universal formula in the
one-dimensional case.  Our formula agrees with that of
Bogomolny-Bohigas-Leboeuf \cite{BBL} and Hannay \cite{H} for $\SU(2)$
polynomials.

\begin{theo} \label{Hformula}
Suppose $\dim M=1$.  Then $$\vec K^N_2(\frac{z}{\sqrtn},\frac{w}{\sqrtn})\to
\vec K^\infty_2(z,w)=\left[\pi\delta_0(z-w)
+H({\textstyle{\half}}|z-w|^2)\right]\frac{ i}{ 2\pi}\ddbar |z|^2 \wedge
\frac{i}{ 2\pi}\ddbar|w|^2\;,$$ where
$$H(t)=
\frac{(\sinh^2 t + t^2) \cosh t  -2t \sinh t }{ \sinh^3 t}=t-\frac{2}{
9}
t^3+\frac{2}{ 45}t^5+O(t^7)\;.$$ \end{theo}

\begin{proof} Making the change of variables $\zeta=z-w$, we have by 
(\ref{UPCC}),
$$\begin{array}{lll}\displaystyle \E\left(\wh Z^N(z) \otimes \wh Z^N(w)\right)
&\to &\displaystyle\frac{i}{ 2\pi}\ddbar|z|^2 \wedge \frac{i}{
2\pi}\ddbar|w|^2-\frac {1}{ \pi^2} \d_z\dbar_z\d_w\dbar_w G(e^{-\half
|z-w|^2})\\[14pt] &&=\ \displaystyle \left[1+{4}\frac{\d^2}{\d z \dbar
z}\frac{\d^2}{\d w \dbar w} G(e^{-\half |z-w|^2}) \right]\frac{i}{
2\pi}\ddbar|z|^2 \wedge\frac {i}{ 2\pi}\ddbar|w|^2\\[14pt] &&=\ \displaystyle
\left[1+{4}\left(\frac{\d^2}{\d\zeta \dbar\zeta}\right)^2 G(e^{-\half
|\zeta|^2}) \right]\frac{i}{ 2\pi}\ddbar|z|^2 \wedge \frac{i}{
2\pi}\ddbar|w|^2\\[14pt] &&=\ \displaystyle \left[1+\frac{1}{ 4}\Delta^2
G(e^{-\half r^2}) \right]\frac{i}{ 2\pi}\ddbar|z|^2 \wedge \frac{i}{
2\pi}\ddbar|w|^2
\end{array}$$
By (\ref{d2}) with $m=1$, we have \begin{eqnarray*}
\Delta^2 G(e^{-\half r^2})
&=&\left(\frac{d^2 }{ dr^2} +\frac{1}{ r}\frac{d }{ dr}\right) 
\left[\log(1-e^{-r^2})+\frac{r^2}{ e^{r^2}-1}\right]\\
&=&4\pi\delta_0 +\frac{8(e^{r^2}-1)^2
-16r^2e^{r^2}(e^{r^2}-1)+4r^4e^{r^2}(e^{r^2}+1)}{ (e^{r^2}-1)^3}\;.
\end{eqnarray*}
Finally,
\begin{eqnarray*} \left[1+\frac{1}{
4}\Delta^2 G(e^{-\half
r^2}) \right] &=& \pi\delta_0 +\frac{(e^{r^2}+1)(e^{r^2}-1)^2
-4r^2e^{r^2}(e^{r^2}-1)+r^4e^{r^2}(e^{r^2}+1)}{ (e^{r^2}-1)^3}\\
&=&\pi\delta_0 +\frac{(\sinh^2\half r^2 +\frac{1}{ 4}r^4)\cosh \half r^2
- r^2
\sinh \half r^2 }{ \sinh^3 \half r^2}\;.
\end{eqnarray*} \end{proof}

\subsection{Pair correlation in higher dimensions}

The limit pair correlation measure is given by
\begin{eqnarray*} \wt K^{\infty}_2(z,w) &=&
\lim_{N\to\infty} N^{2(m-1)}\wt K^N_2(\frac{z}{\sqrtn},\frac{w}{\sqrtn})\\ & =
& \vec K^{\infty}_2(z,w)\wedge \frac{1}{(m-1)!}\left(\frac{i}{2}\ddbar
|z|^2\right)^{m-1}\wedge \frac{1}{(m-1)!}\left(\frac{i}{2}\ddbar
|w|^2\right)^{m-1}\,.\end{eqnarray*} (The scaling $N^{2(m-1)}$ comes from the
fact that $N\om(\frac{z}{\sqrtn})
=N(\tau_{\sqrt{N}})_*\om\to
\frac{i}{2}\ddbar |z|^2$.)
We now compute $\wt K^{\infty}_2$
for the case of a
manifold of general dimension
$m>1$. It is convenient to express this measure in terms of the
expected density of zeros 
\begin{equation}\label{expden} \wt K^\infty_1(z)=
\lim_{N\to\infty}N^{m-1}\wt K^N_1(\frac{z}{\sqrtn})=
\frac{m}{\pi}dV_{\C^m}
=\frac{1}{\pi(m-1)!}\left(\frac{i}{2}\ddbar|z|^2\right)^m\,. \end{equation}
We have the following explicit universal formula for the limit pair
correlation measure.  In particular, it gives the scaling limit pair
correlation for the zeros of $\SU(m+1)$-polynomials.

\begin{theo} \label{mformula} Suppose $\dim M=m>1$.  Then
$$\wt K^\infty_2(z,w)=
\left[\gamma_m({\textstyle{\half}}|z-w|^2)\right]
\wt K^\infty_1(z)\wedge \wt K^\infty_1(w)
\;,$$
where \begin{eqnarray*}\gamma_m(t)&=& 
\frac{\left[{\textstyle\half}(m^2+m)\sinh^2t + t^2\right]\cosh t-(m+1)t
\sinh t }{ m^2 \sinh^3 t} +\frac{m-1}{ 2m}\\
&=&\frac{(m-1)}{ 2m}t^{-1}+\frac{m-1}{ 2m}+\frac{(m+2)(m+1)}{
6m^2}t\\
&&\quad\quad -\frac{(m+4)(m+3)}{
90m^2}t^3+
{\frac {\left (m+6\right )\left (m+5\right )}
{945{m}^{2}}}{t}^{5} +O(t^7) 
\;.\end{eqnarray*}\end{theo}

\begin{proof} By (\ref{UPCC}) and
(\ref{d2}), again writing $\zeta=z-w$ (except this time $\zeta\in\C^m$),
\begin{eqnarray}&&\wt K^\infty_2(z,w)\ = \ 
\left[1+\frac{4}{m^2}\sum_{j,k=1}^m \frac{\partial^2}{\partial z_j
\partial \bar{z}_j } \frac{\partial^2}{\partial w_k \partial \bar{w}_k }
G(e^{-\half|z-w|^2})\right]  
\wt K^\infty_1(z)\wedge \wt K^\infty_1(w)\nonumber \\ &&\qquad =\ \left[1+
\frac{1}{4 m^2}
\Delta^2_\zeta G(e^{-\half|\zeta|^2})\right] \wt K^\infty_1(z)\wedge \wt
K^\infty_1(w)\nonumber\\
&&\qquad  =\ \left[ 1 + \frac{1}{4m^2}
\left(\frac{d^2}{dr^2} +
\frac{2m -1}{r} \frac{d}{dr}\right)\left( m \log(1 - e^{-r^2}) +
\frac{r^2}{e^{r^2} - 1}\right)\right]\wt K^\infty_1(z)\wedge \wt
K^\infty_1(w)\;.\label{pc-m}
\end{eqnarray}  Computing the Laplacian in (\ref{pc-m}) leads to the stated
formula.  \end{proof}

Note that if we substitute $m=1$ in the expression for $\gamma_m(t)$, we
obtain Hannay's function $H(t)$.  However for the case $m>1$, the limit
measure is absolutely continuous on $\C^m\times \C^m$, whereas in the
one-dimensional case, there is a self-correlation  delta measure.

\end{document}